\documentclass[aps,prl,twocolumn,superscriptaddress,10pt]{revtex4-2}
\usepackage{multirow,amsthm,amssymb,amsbsy,amsmath,epsfig,epstopdf,bm,float}
\DeclareMathOperator{\Tr}{Tr}
\usepackage{enumitem}
\usepackage{graphicx}
\usepackage{booktabs}
\usepackage{verbatim}
\usepackage{dcolumn}
\usepackage{color}
\usepackage[dvipsnames]{xcolor}
\usepackage{mathtools}
\usepackage[normalem]{ulem}
\usepackage{soul}
\usepackage{hyperref}
\usepackage{braket}
\usepackage[export]{adjustbox}
\DeclarePairedDelimiter\abs{\lvert}{\rvert}%

\DeclarePairedDelimiter\norm{\lVert}{\rVert}

\begin{document}

\title{Collision models can efficiently simulate any multipartite Markovian quantum dynamics
}

\author{Marco Cattaneo}
\email{marcocattaneo@ifisc.uib-csic.es }
\affiliation{Instituto de F\'{i}sica Interdisciplinar y Sistemas Complejos (IFISC, UIB-CSIC), Campus Universitat de les Illes Balears E-07122, Palma de Mallorca, Spain}
\affiliation{QTF Centre of Excellence, Turku Centre for Quantum Physics, 
Department of
Physics and Astronomy, University of Turku, FI-20014 Turun Yliopisto, Finland}
\affiliation{QTF Centre of Excellence,  
Department of Physics, University of Helsinki, P.O. Box 43, FI-00014 Helsinki, Finland}

\author{Gabriele De Chiara}
\affiliation{Centre for Theoretical Atomic, Molecular and Optical Physics, Queen’s University Belfast, Belfast BT7 1NN, United Kingdom}

\author{Sabrina Maniscalco}
\affiliation{QTF Centre of Excellence,  
Department of Physics, University of Helsinki, P.O. Box 43, FI-00014 Helsinki, Finland}
\affiliation{QTF Centre of Excellence, Turku Centre for Quantum Physics, 
Department of
Physics and Astronomy, University of Turku, FI-20014 Turun Yliopisto, Finland}
\affiliation{QTF Centre of Excellence, Department of Applied Physics, School of 
Science, Aalto University, FI-00076 Aalto, Finland}

\author{Roberta Zambrini}
\affiliation{Instituto de F\'{i}sica Interdisciplinar y Sistemas Complejos (IFISC, UIB-CSIC), Campus Universitat de les Illes Balears E-07122, Palma de Mallorca, Spain}

\author{Gian Luca Giorgi}
\affiliation{Instituto de F\'{i}sica Interdisciplinar y Sistemas Complejos (IFISC, UIB-CSIC), Campus Universitat de les Illes Balears E-07122, Palma de Mallorca, Spain}

\date{Received: \today / Accepted: date}

\begin{abstract}
%We introduce the \textit{multipartite collision model} to simulate the Markovian dynamics of any multipartite open quantum system by decomposing the system-environment interaction into elementary collisions between subsystems and ancillas, thus providing a simple decomposition in terms of elementary quantum gates for quantum computation. The generality of the model allows for the study of any possible Markovian global and local master equation in the presence of any kind of bath at any temperature, and makes it particularly suitable to address scenarios in which the fundamental interaction and energy exchange between subsystems and environment takes on great importance, e.g. in quantum thermodynamics. Moreover, we develop a method to estimate an analytical error bound for any repeated interactions model, and we use it to show that the error of the \textit{multipartite collision model} displays an optimal behavior. Finally, we proof that the \textit{multipartite collision model} is efficiently simulable on a quantum computer according to the dissipative quantum Church-Turing theorem.  

We introduce the \textit{multipartite collision model}, defined in terms of elementary interactions between subsystems and ancillas, and show that it can simulate the Markovian dynamics of any multipartite open quantum system.  {We} develop a method to estimate an analytical error bound for any repeated interactions model, and we use it to prove that the error of our scheme displays an optimal {scaling}. Finally, we provide a simple decomposition of the \textit{multipartite collision model} into elementary quantum gates, and show that it is efficiently simulable on a quantum computer according to the dissipative quantum Church-Turing theorem, i.e. it requires a polynomial number of resources.
\end{abstract}
\maketitle

\textit{Introduction}.--The collision approach represents one of the most successful methods to describe the dynamics of an open quantum system, being based on the intriguing idea that enviroment-induced decoherence and dissipation arise because of rapid repeated collisions between each system unit and a set of environment ancillas, occurring during a timestep $\Delta t$. This framework, whose origins can be traced back to some important works of the previous century \cite{Karplus1948,Rau1963,Dumcke1985}, has given birth to a pletora of ``collision'' or ``repeated interactions'' models \cite{Scarani2002,Ziman2002,Ziman2005,Attal2006a,Giovannetti2012,Ciccarello2013a,Bruneau2014a,Vacchini2014,Grimmer2016a,Kretschmer2016a}, which have been receiving an increasing attention in recent years, especially due to their fundamental importance in the fields of quantum thermodynamics and open quantum systems.
For instance, collision models have been proven useful to investigate flux rectification \cite{Landi2014}, Landauer's principle \cite{Lorenzo2015a,PezzuttoNJP2016}, the emergence of thermalization or non-equilibrium steady states \cite{Karevski2009,Barra2017,Lostaglio2018,Cusumano2018a,Seah2019a,Arisoy2019,Manatuly2019a,Korzekwa2020,Ehrich2020,Guarnieri2020a}, quantum thermometry \cite{Seah2019}, quantum batteries \cite{Barra2019} and quantum thermal machines \cite{Dag2016,Pezzutto2019,Hewgill2020,DeChiara2020,Piccione2020}, as well as to analyze the thermodynamics of non-thermal baths \cite{Manzano2016,Manzano2018,Rodrigues2019} or in the presence of strong coupling \cite{Strasberg2019a}. 
Applications outside the field of thermodynamics include the study of open quantum optical systems \cite{Bruneau2014,Ciccarello2017,Cilluffo2020,Carollo2020}, simulation of non-Markovian effects \cite{Ciccarello2013a,Bernardes2014,McCloskey2014,Vacchini2014,Kretschmer2016a,Cakmak2017a,Lorenzo2017,Campbell2018,Jin2018a,Man2018a} and cascade models \cite{Giovannetti2012,Lorenzo2015a,Lorenzo2015,Cusumano2017a,Cusumano2018}, quantum synchronization \cite{Karpat2019,karpat2020synchronization}, entanglement generation \cite{Daryanoosh2018,Cakmak2019,Li_2020}, quantum transport \cite{Chisholm2020} and quantum Darwinism \cite{Campbell2019a,Garcia-Perez2020}.

The structure of any single-qubit collision model and {the correspondence with an equivalent} master equation is well-understood \cite{Ziman2005,Rybar2012,Filippov2017}. In contrast, while some collision models for multipartite systems have been presented in the past few years \cite{Daryanoosh2018,Lorenzo2017,Giovannetti2012,Supplemental}, a universal protocol suitable for efficient simulation of multipartite open system dynamics via collision models, described in terms of elementary collisions between subsystems and ancillas, has not been provided yet. Reproducing any possible open dynamics by means of elementary collision models promises to be particularly valuable to deal with the microscopic description of multipartite open systems, where \textit{global} master equations are needed \cite{Hofer2017,Gonzalez2017,Cattaneo2019b} and one cannot always rely on local descriptions, that may display fundamental differences e.g. from the thermodynamic point of view \cite{Levy2014,Stockburger2017}. Here, collision models are extremely useful to study the elementary exchange of heat and energy, and the microscopic production of work in each single interaction between a unit of the system and an environment ancilla \cite{Barra2015,Strasberg2017,DeChiara2018a}. For instance, a collision model analysis resolves the violation of the second law of thermodynamics when using a local master equation \cite{DeChiara2018a,Hewgill2020a}.

In this Letter, we introduce the \textit{multipartite collision model} (MCM), 
based on elementary interactions between each unit of a multipartite system and a set of ancillary qubits of the environment. We show that the MCM is able to reproduce any Gorini-Kossakowski-Sudarshan-Lindblad (GKLS) master equation \cite{Lindblad1976,Gorini1976a} in the limit of small timestep $\Delta t\rightarrow 0^+$, therefore describing any possible divisible dynamical map. {After providing a simple decomposition into elementary quantum gates, we prove that the MCM is efficiently simulable on a quantum computer under the assumptions of the dissipative quantum Church-Turing theorem \cite{Kliesch2011}, as it requires a number of resources that scales polynomially as a function of the number of subsystems, time, and the inverse of desired precision.} This allows for the efficient simulation of a whole range of complex open quantum systems under the Markovianity assumption: by tuning our model in an intuitive way, we can mimic the effect of different types of separate and/or common baths (bosonic, fermionic, spin, etc.) at any temperature, as well as reproduce each elementary system-bath interaction characterizing a generic global master equation, with or without a non-local unitary system dynamics. Non-Markovian effects may be then simulated by Markovian embeddings of pseudomodes into the MCM \cite{Lorenzo2017}. Furthermore, by developing a method valid for \textit{any} collision model, we calculate an analytical error bound for the simulation of a generic semigroup dynamics by means of the MCM, proving that its scaling is optimal.

To guarantee the generality of the MCM, we {will show that it can simulate the dynamics driven by} any GKLS master equation, both in the diagonal \cite{Lindblad1976} and non-diagonal form \cite{Gorini1976a}. The latter can be expressed by means of the Liouvillian superoperator $\mathcal{L}$ as:
\begin{equation}
\label{eqn:LindbladNoDiag}
\begin{split}
\mathcal{L}[\rho_S(t)]=&-i[\tilde{H}_S,\rho_S(t)]+\sum_{j,k=1}^{J} \gamma_{jk}\mathcal{D}_{F_j,F_k}[\rho_S(t)],
\end{split}
\end{equation}
where $\mathcal{D}_{O_1,O_2}[\rho]=O_1\rho O_2^\dagger-\frac{1}{2}\{O_2^\dagger O_1,\rho\}$, {$\tilde{H}_S$} is an effective {system} Hamiltonian, $\gamma_{jk}$ is the semipositive \textit{Kossakowski matrix}, while we term $\{F_k\}_{k=1}^J$ as \textit{GKS operators} {\cite{Gorini1976a}}. If $\mathcal{H}_S=\bigotimes_{j=1}^M \mathcal{H}_S^{(j)}$ is the Hilbert space of the system composed of (for simplicity identical) $M$ subsystems with ${\rm dim}(\mathcal{H}_S^{(j)})=d$ ($d<\infty$), in general we have $J=d^{2M}-1$. We obtain the diagonal GKLS form by diagonalizing the Kossakowski matrix through a suitable unitary matrix $C$: we introduce the \textit{Lindblad operators} {\cite{Lindblad1976}} $L_k=\sum_{j=1}^J C_{jk} F_j$, and we derive the corresponding decay rates $\Gamma_k$ as the eigenvalues of $\gamma_{jk}$.

\begin{figure}
\includegraphics[scale=0.3]{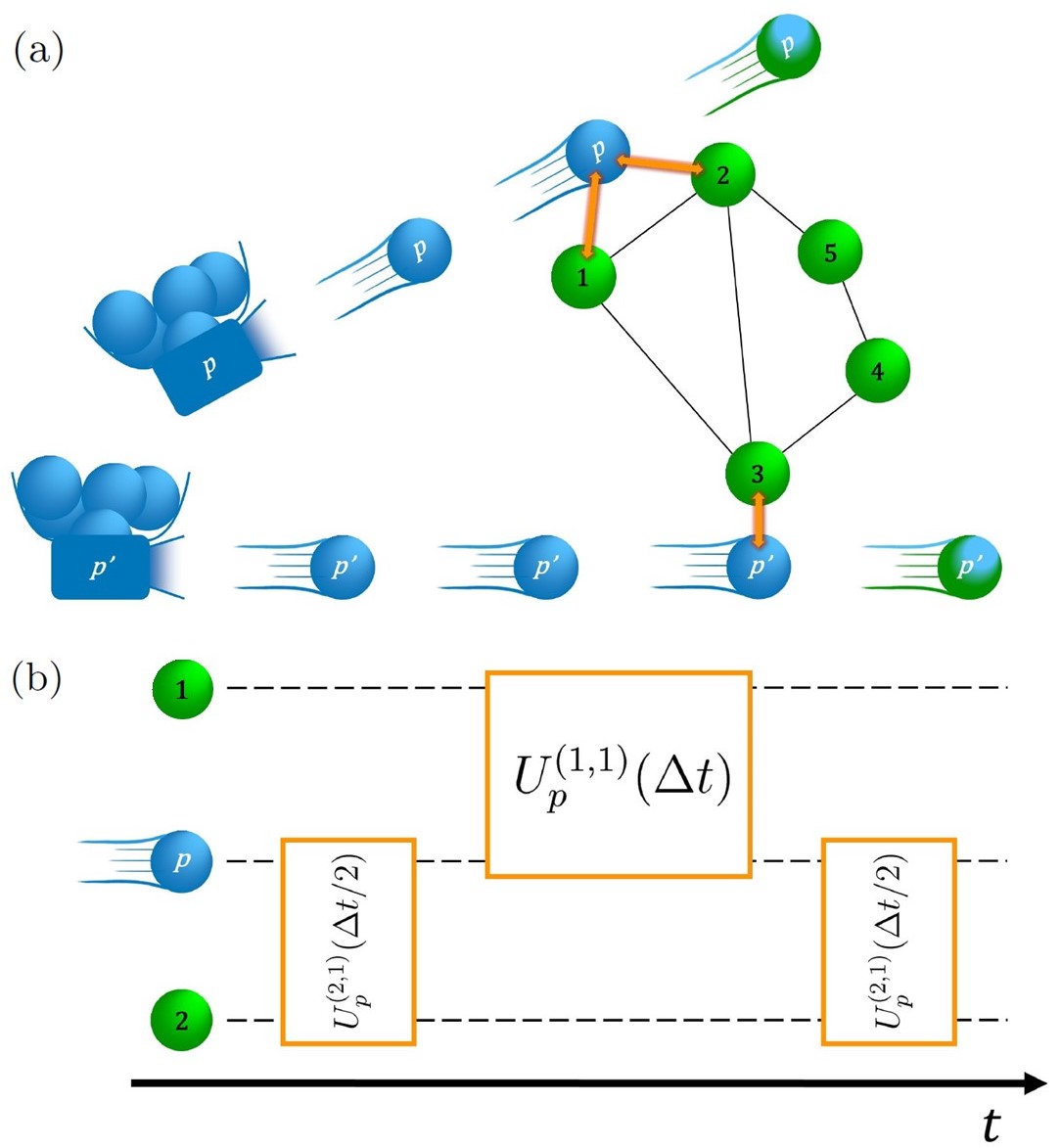}
\caption{(a): Pictorial representation of the MCM. The ancilla $p$ generates a term in the master equation that couples subsystems $1$ and $2$. The ancilla $p'$ interacts with subsystem $3$ only, and yields a local term in the master equation. (b): Circuit scheme of the interaction between ancilla $p$ and subsystems $1$ and $2$. If the system is made of qubits, three two-qubit gates are required.}
\label{fig:collMod}
\end{figure}
For the sake of clarity, we begin by assuming that each GKS operator $F_k$ {acts} non-trivially on a single subsystem only, although the MCM is not restricted to it, {as we will see in the following}. {This assumption is satisfied by a wide range of local and global master equations \cite{Cattaneo2019b}, and corresponds to neglecting environment-mediated many-body interactions between the subsystems.} Under this assumption, the total number of Lindblad operators reduces to $J=M(d^2-1)$, and the index $j=(m,\alpha)$ {can be decomposed into} two additional indexes: $m=1,\ldots,M$ labeling the subsystems and $\alpha=1,\ldots,(d^2-1)$ selecting the specific GKS operator acting locally thereon.

\textit{\textit{Multipartite collision model}}.--The procedure to implement the MCM under the assumption of local GKS operators is depicted in Fig.~\ref{fig:collMod}. For the non-diagonal case we can identify the following five steps: 
\begin{enumerate}[wide, labelwidth=!, labelindent=0pt]
\item For each pair of GKS operators $F_{m,\alpha}$ and $F_{m',\alpha'}$ appearing in Eq.~\eqref{eqn:LindbladNoDiag}, consider an independent ancillary qubit of the environment labeled by {$p=(m,\alpha,m',\alpha')$}, and construct the sequence of local elementary subsystem-ancilla interactions given by:
\begin{equation}
\label{eqn:pairEv}
U_p(\Delta t)=U_p^{(m,\alpha)}(\Delta t/2)U_p^{(m',\alpha')}(\Delta t)U_p^{(m,\alpha)}(\Delta t/2),
\end{equation}
where {($\hbar=1$)}
\begin{equation}
\label{eqn:elemInt}
U_p^{(m,\alpha)}(\Delta t)=\exp(-i g_I \Delta t H_{I,p}^{(m,\alpha)}).
\end{equation}
$H_{I,p}^{(m,\alpha)}=(\lambda_p^{(m,\alpha)}F_{m,\alpha}\sigma_{p}^++h.c.)$, $g_I$ is a fixed constant with the units of energy and $\lambda_p^{(m,\alpha)}$ is a dimensionless parameter we can freely tune.
\item Compose all the unitary evolutions associated to each pair of GKS operators into a global unitary operator describing the overall interaction with the environment, choosing freely the order in which we insert the former:
\begin{equation}
\label{eqn:evComposition}
U_I(\Delta t)=\prod_{p\in \mathsf{P}}U_p(\Delta t),
\end{equation}
where the elements of the set $\mathsf{P}$ are all the possible pairs $(m,\alpha,m',\alpha')$.
\item Add a unitary system evolution driven by the dimensionless system Hamiltonian $H_S$ to obtain the final global operator for the simulation of the MCM:
\begin{equation}
\label{eqn:evSimTot}
U_{sim}(\Delta t)=U_S(\Delta t)\circ U_I(\Delta t),
\end{equation}
with $U_S(\Delta t)=\exp(-i g_S\Delta t H_S)$, where $H_S=\tilde{H}_S/g_S$ and $g_S$ is a fixed constant with the units of energy, defining the order of magnitude of $\tilde{H}_S$.
\item Prepare the set of environment {qubits} with $p\in\mathsf{P}$ in an initial separable state $\rho_E(0)=\bigotimes_{p\in\mathsf{P}} \eta_p$, where $\eta_p=c_p\ket{\downarrow}_p\bra{\downarrow}+(1-c_p)\ket{\uparrow}_p\bra{\uparrow}$, with $0\leq c_p\leq 1$, is a diagonal state in the basis of $\sigma_{p}^z$.
\item Apply a single step of the MCM on the system state $\rho_S$ as the quantum map:
\begin{equation}
\label{eqn:quantumMap}
\phi_{\Delta t}[\rho_S]=\Tr_E\left[U_{sim}(\Delta t)\rho_S\otimes\rho_E(0)U_{sim}^\dagger(\Delta t)\right],
\end{equation}
where the trace over the environment $E$ includes the trace over each environment ancilla with $p\in\mathsf{P}$.
%\item Reconstruct the open system evolution at time $t=n\Delta t$, with $n\in \mathbb{N}$, as a repeated application of the quantum map:
%\begin{equation}
%\rho_S(t)=\left(\phi_{\Delta t}\right)^n[\rho_S(0)].
%\end{equation} 
\end{enumerate}

We show in the Supplemental Material \cite{Supplemental} that under certain requirements the dynamics generated by the MCM corresponds to the one driven by a general GKLS master equation~\eqref{eqn:LindbladNoDiag}. Specifically, we follow the standard derivation of a collision model \cite{Lorenzo2017}: we assume the limit of small timestep $\Delta t\rightarrow 0^+$, with 
$g_S\ll g_I \ll \Delta t^{-1}$, and $\lim_{\Delta t\rightarrow 0^+} g_I^2\Delta t=\gamma$,
where $\gamma$ is a finite energy constant. 
For simplicity, the coefficients $\lambda_p^{(m,\alpha)}$ in Eq.~\eqref{eqn:elemInt} are taken of the order of $O(1)$.
Under the above assumptions, the evolution generated by a single application of the quantum map corresponds to:
\begin{equation}
\label{eqn:approxMap}
\phi_{\Delta t}=\mathbb{I}+\Delta t \mathcal{L}+O(\Delta t^2),
\end{equation}
where the Liouvillian superoperator reads \cite{Supplemental}:
\begin{equation}
\label{eqn:Liouvillian}
\begin{split}
\mathcal{L}[\rho_S]=&-i[\tilde{H}_S,\rho_S]+\sum_{p\in\mathsf{P}}\left(\gamma_p^\downarrow    \mathcal{D}_{F_{m,\alpha},F_{m',\alpha'}}[\rho_S] \right.\\
&+\left. \gamma_p^\uparrow \mathcal{D}_{F_{m,\alpha}^\dagger,F_{m',\alpha'}^\dagger}[\rho_S] +h.c.\right).
\end{split}
\end{equation}
The coefficients are:
\begin{equation}
\label{eqn:coefficients}
\begin{split}
&\gamma_p^\downarrow=\begin{cases}
\gamma c_p \lambda_p^{(m,\alpha)}(\lambda_p^{(m',\alpha')})^*&\textnormal{ if }m\neq m'\textnormal{ or }\alpha\neq\alpha'\\
\gamma\sum_{\bar{p}} c_{\bar{p}} \abs{\lambda_{\bar{p}}^{(m,\alpha)}}^2 &\textnormal{ otherwise}\\
\end{cases}\\
&\gamma_p^\uparrow=\begin{cases}
\gamma(1-c_p) (\lambda_p^{(m,\alpha)})^*\lambda_p^{(m',\alpha')}&\textnormal{ if }m\neq m'\textnormal{ or }\alpha\neq\alpha'\\
\gamma\sum_{\bar{p}} (1-c_{\bar{p}}) \abs{\lambda_{\bar{p}}^{(m,\alpha)}}^2 &\textnormal{ otherwise},\\
\end{cases}\\
\end{split}
\end{equation}
with summation over all the unordered pairs of GKS operators  $\bar{p}=(m,\alpha,\bar{m},\bar{\alpha})$. These coefficients give rise to a semipositive Kossakowski matrix, i.e. the master equation~\eqref{eqn:Liouvillian} is already in GKLS form. {Eq.~\eqref{eqn:Liouvillian} also contains all the terms associated to the adjoint GKS operators with Kossakowski matrix $\gamma_p^{\uparrow}$, that can be removed by setting $c_p=1$ $\forall p$ (i.e. by preparing each ancilla in the ground state).} 
Given the freedom in the choice of $\lambda_{p}^{(m,\alpha)}$ and $\lambda_{p}^{(m',\alpha')}$ in the Hamiltonian of each elementary subsystem-ancilla interaction introduced in Eq.~\eqref{eqn:elemInt}, we can engineer $\gamma_p^{\downarrow}$ in order to reproduce any Kossakowski matrix for the GKS operators of the master equation~\eqref{eqn:Liouvillian}, and therefore any non-diagonal GKLS master equation~\eqref{eqn:LindbladNoDiag} {with effective Hamiltonian $\tilde{H}_S$}. We can therefore conclude that repeated rapid applications of the MCM simulate the quantum semigroup dynamics driven by any Liouvillian $\mathcal{L}$:
\begin{equation}
\label{eqn:semigroupEv}
\lim_{\Delta t\rightarrow 0^+} \left(\phi_{\Delta t}\right)^n=\exp\mathcal{L}t,\textnormal{ with }t=n\Delta t.
\end{equation}
{This is our first major result.}
For a small but finite $\Delta t$, the MCM reproduces the open dynamics only for discrete times $t=n\Delta t$, where the resolution given by $\Delta t$ can be thought of as the \textit{coarse-graining} of the master equation \cite{Farina2019}. Finally, it is not always necessary to take one ancilla for each pair of jump operators. In certain scenarios we may rely on a simpler version of the MCM that requires a smaller number of resources \cite{Supplemental}.

The collision scheme introduced above is particularly useful in situations where one has to apply the MCM to a symbolic GKLS master equation that cannot be diagonalized analytically. In all other cases, the MCM realizes the diagonal form of the GKLS master equation by following the same lines described above, with the prescription that we just need one ancillary qubit for each Lindblad operator $L_k$. Indeed, under the assumption of local GKS operators, we can write $L_k=\sum_{m=1}^M \tilde{F}_{m}^{(k)}$, {where $\tilde{F}_{m}^{(k)}=\sum_{\alpha=1}^{d^2-1} C_{\alpha k} F_{m,\alpha}$ is a local sum of GKS operators}, and the evolution in Eq.~\eqref{eqn:pairEv} is replaced by the sequence of elementary interactions 
\begin{equation}
\label{eqn:diagU}
U_k(\Delta t)=\prod_{m=1}^M U_k^{(M-m+1)}(\Delta t/2)\prod_{m'=1}^M U_k^{(m')}(\Delta t/2),
\end{equation}
and $U_k^{(m)}(\Delta t)=\exp [-i g_I \Delta t (\lambda_k \tilde{F}_m^{(k)}\sigma_k^+ + h.c.)] $, so that $\Gamma_k=\lim_{\Delta t\rightarrow 0^+} g_I^2 \Delta t \abs{\lambda_k}^2 $ is the decay rate of the $k$th Lindblad operator. Correspondingly, the product in the global unitary operator for the interaction with the environment in Eq.~\eqref{eqn:evComposition} runs over $k=1,\ldots,J$ instead of the pairs $p\in\mathsf{P}$.

\textit{Temperature}.--Note that a suitable engineering of the parameters of the MCM allows for the simulation of any thermal bath at any (even negative) temperature. For instance, to mimic a single thermal bath at temperature $T$, one can use a single ancilla prepared in a thermal state at temperature $T$, and the strength of the decay rates can be engineered by tuning the parameters $\lambda_p^{(m,\alpha)}$ as a function of $T$ \cite{Supplemental}. Our model also allows for energy-nonconserving elementary interactions (e.g. with counter-rotating terms such as $a^\dagger\sigma_p^++h.c.$ for a bosonic mode $a$ and a qubit ancilla labeled by $p$). This may generate squeezing-like terms,
which corresponds to the ancillas not having the same temperature as the effective bath, and any complex scenario with multiple baths can be realized. This engineering overcomes the physical constraints of previous open system quantum simulations based on qubit ancillas \cite{Wang2011}.

\textit{Extension to many-body GKS operators and time-dependent semigroups}.--The MCM also works in the case of many-body GKS operators that cannot be trivially decomposed into single subsystem-ancilla interactions {(e.g. when a GKS operator is written as $F_j=\sigma_1^-\sigma_2^+$ \cite{Manzano2019})}. In this scenario, we will have an elementary collision in Eq.~\eqref{eqn:elemInt} with Hamiltonian $H_{I,p}^{(j)}=\lambda_p^{(j)}F_j\sigma_{p}^++h.c.$, where $F_j$ acts non-trivially on more than one subsystem, and therefore cannot be represented by a single two-qubit gate on a quantum computer. Its action may be implemented by multi-qubit gates, as already done in quantum simulation of open systems \cite{Barreiro2011}, or by a decomposition in terms of two-qubit gates \cite{nielsenchuang}. In general terms, we may assume to have at our disposal a set of $R$ Hamiltonians for each GKS operator $F_j$ (or Lindblad operator $L_j$ in the diagonal case), $H_r^{(j)}=G_r^{(j)}\sigma_{p}^++h.c.$ ($p$ labels a generic ancilla), that we are able to simulate by elementary multi-qubit gates in our lab, through which we can build any required GKS operator Hamiltonian in Eq.~\eqref{eqn:elemInt} as $H_{I,p}^{(j)}=\sum_{r=1}^R\mu_r^{(p,j)} H_r^{(j)}$. Then, we can simulate the MCM by the decomposition $\exp(-ig_I\Delta tH_{I,p}^{(j)})=\prod_{r=1}^R U_{p,R-r+1}^{(j)}(\Delta t/2)\prod_{r'=1}^R U_{p,r'}^{(j)}(\Delta t/2)+O(g_I^3\Delta t^3)$, with $U_{p,r}^{(j)}(\Delta t)=\exp(-ig_I\Delta t\mu_r^{(p,j)} H_r^{(j)})$, which still brings an error of the order of $O(\Delta t^2)$ in Eq.~\eqref{eqn:approxMap} \cite{Supplemental}. {Note that, if we go back to the condition of local GKS operators,} for simplicity we can assume to be able to directly implement any elementary gate $U_{p}^{(m,\alpha)}$ in the lab, and therefore for the non-diagonal case $R=1$. In the diagonal case, we can interpret $R$ as the number of different elementary subsystem-ancilla interactions in Eq.~\eqref{eqn:diagU}, therefore $R=M$. Finally, extensions to time-dependent semigroups in which the Kossakowski matrix in Eq.~\eqref{eqn:LindbladNoDiag} depends on time, $\gamma_{jk}(t)$ semipositive for any time $t$, are {immediate}: we just need to set a time-dependent parameter $\lambda_p^{(m,\alpha)}(t)$ in the Hamiltonian of Eq.~\eqref{eqn:elemInt}, and to make it vary as a function of $t$. Analogously, we can make the system Hamiltonian depend on time as well, as $H_S(t)$.

\textit{Error estimation}.--Previous treatments of the error analysis for a collision model have usually neglected higher-order terms in the Taylor expansion, e.g. see the detailed discussion in Ref.~\cite{Grimmer2016a}. Sometimes this may not be accurate, since the infinite series of higher-order terms may bring a non-negligible contribution \cite{Childs2019}. Here, we estimate an analytical error bound for the MCM by keeping all the terms of the infinite Taylor expansion through a method based on Suzuki's higher-order integrators \cite{Suzuki1985} which can be found in the Supplemental Material \cite{Supplemental}, whose validity applies to any collision model. For the sake of a general description, we compute the error bound without assuming the GKS operators locality, and therefore relying on the sets of $R$ many-body Hamiltonians $H_r^{(j)}$ introduced above.

To estimate the error bound we employ the $1\rightarrow 1$ superoperator norm $\norm{\mathcal{T}}_{1\rightarrow 1}$ and the operator norm $\norm{A}_\infty$ \footnote{The superoperator norm is defined as \cite{Kliesch2011,Sweke2015} $\norm{\mathcal{T}}_{1\rightarrow 1}=\sup_{\norm{A}_1=1}\norm{\mathcal{T}[A]}_1$,
where $\norm{A}_1=\Tr[\sqrt{A^\dagger A}]$ is the trace norm. The operator or infinity norm is defined as $\norm{A}_\infty=\sup_{\norm{v}=1}\norm{Av}$, where $\norm{v}$ is the standard vector norm.}. We can identify four different kinds of error made by approximating the semigroup evolution through the collision model:
\begin{description}
\item[Global error] $\epsilon_g=\norm{\exp\mathcal{L}t-(\phi_{\Delta t})^n}_{1\rightarrow 1}$, with $\Delta t=t/n$.
\item[Single-step error] $\epsilon_s=\norm{\exp\mathcal{L}\Delta t-\phi_{\Delta t}}_{1\rightarrow 1}$.
\item[Truncation error] $\epsilon_t=\norm{\exp\mathcal{L}\Delta t-(\mathbb{I}+\Delta t\mathcal{L})}_{1\rightarrow 1}$.
\item[Collision error] $\epsilon_c=\norm{\phi_{\Delta t}-(\mathbb{I}+\Delta t\mathcal{L})}_{1\rightarrow 1}$.
\end{description}
Following \textit{Lemma 2} in Ref.~\cite{Sweke2014}, we have $\epsilon_g\leq n\epsilon_s$, and according to the triangle inequality $\epsilon_s\leq\epsilon_t+\epsilon_c$. The latter errors can be bound as \cite{Supplemental}:
\begin{equation}
\label{eqn:boundT}
\epsilon_t\leq 2e (R\Lambda(1+J R\Lambda)\Delta t)^2,\;2R\Lambda(1+J R\Lambda)\Delta t<1,
\end{equation}
\begin{equation}
\label{eqn:boundC}
\epsilon_c\leq \textnormal{pol}_1(\Lambda,\Xi,g_S,\gamma)\Delta t^2+\textnormal{pol}_2(\Lambda,\Xi,g_S,\gamma)\Delta t^3,
\end{equation}
where $\textnormal{pol}_1$ and $\textnormal{pol}_2$ are polynomial functions of $g_S,\gamma,\Lambda=\max_{r,j,p}(\norm{H_S}_\infty,\norm{\mu_r^{(p,j)} H_r^{(j)}}_\infty)$ and $\Xi$, equal to the total number of different elementary unitary evolutions driven by a single $H_r^{(j)}$ in Eq.~\eqref{eqn:evComposition} (in the case of MCM for the diagonal master equation, we have $\Xi=R J$, for the non-diagonal scenario $\Xi=R \abs{\mathsf{P}}$). The exact expressions of $\textnormal{pol}_1$ and $\textnormal{pol}_2$, as well as the above bounds under the assumption of $k$-locality \cite{Kliesch2011},  {are discussed with further details} in the Supplemental Material \cite{Supplemental}. Here, we just remark that the global error of the MCM follows the behavior:
\begin{equation}
\label{eqn:globalErrBeh}
\epsilon_g=O(n\Delta t^2)=O(t^2/n).
\end{equation}
This {scaling is optimal} for the error made by simulating an open system dynamics via a general scheme of repeated unitary evolutions \cite{Cleve2017}, and therefore via general collision models. Such scaling, for instance, is always saturated by the truncation error $\epsilon_t$, which is the same for any model of rapid repeated interactions. {This is our second major result}.

\textit{Resource estimation for quantum simulation}.--To address the quantum simulation efficiency of the MCM we assume the $k$-locality of the Liouvillian $\mathcal{L}$, {namely, that} it can be written as a sum of Liouvillians $\mathcal{L}_\sigma$ non-trivially acting on $k$ subsystems only: $\mathcal{L}=\sum_{\sigma=1}^K\mathcal{L}_\sigma$. This is a standard assumption for quantum simulation on a circuital quantum computer, introduced by Kliesch et al. for open systems \cite{Kliesch2011,Barthel2012}, and first imposed in the seminal paper by Lloyd on Hamiltonian quantum simulation \cite{Lloyd1996}. $K\leq M^k$ is the total number of possible $k$-local terms, that for {large} $M$ goes as $K\sim M^k/(k!e^k)$ \cite{Supplemental}. We estimate the number of resources focusing on the MCM for the diagonal GKLS master equation only, given that this is certainly the most convenient scheme for the simulation on a quantum computer. We allow for many-body GKS operators, and we count the number of elementary gates driven by the sets of $R$ Hamiltonians $\{H_r^{(\sigma,j)}\}_{r=1}^R$, corresponding to the Lindblad operators $L_{\sigma,j}$ of {each} $\mathcal{L}_\sigma$. {Note that $k$-locality implies $R< d^{2k}$.} Under these assumptions and with $H_S=\sum_{\sigma=1}^K H_S^{(\sigma)}$, the error bound in Eq.~\eqref{eqn:boundC} is conveniently rewritten by substituting $\Lambda\rightarrow\Lambda'=\max_{r,j,\sigma}(\norm{H_S^{(\sigma)}}_\infty,\norm{\mu_r^{(\sigma,j)} H_r^{(\sigma,j)}}_\infty)$, $\Xi\rightarrow \Xi'=K R J_k$ \cite{Supplemental}, where the total number of Lindblad operators for $k$-local Liouvillians is bound by $J_k\leq d^{2k}-1$. $\Lambda'$ does not increase with the total number of subsystems, while $\Xi'$ scales polynomially with $M$. Moreover, the bound in Eq.~\eqref{eqn:boundT} is multiplied by $K^2$ and modified with $\Lambda\rightarrow\Lambda'$, $J\rightarrow J_k$ as above, thus it scales polynomially with $M$. Therefore, we can set $\epsilon_g\leq f(M)t^2/n$, where $f(M)$ is a polynomial function of the total number of subsystems \cite{Supplemental}.

For a single timestep of the MCM, we need one ancilla for each Lindblad operator of each $k$-local Liouvillian. Therefore, we require $K J_k$ ancillas per timestep. For the simulation up to time $t$ within a global precision of $\epsilon_g$, we need $
N_A=\left\lceil K\cdot J_k\cdot f(M)t^2/\epsilon_g\right\rceil$ ancillas,
which is a polynomial function $\textnormal{poly}(M,t,1/\epsilon_g)$ and therefore provides us with an efficient number of ancillas \footnote{We are assuming to use a new set of ancillas at each timestep. In case of not having at our disposal a large number of available ancillas, one may also reinitialize the set of $K J_k$ ancillas to the state $\rho_E(0)$ before each timestep, and the results of this work would still hold.} for quantum simulation \cite{Kliesch2011,Sweke2015}.

We need $2R-1$ elementary quantum gates for each Lindblad operator of a single timestep. Hence, the total number of gates in a single timestep is $(2R-1) K J_k+N_G^{(S)}$, where $N_G^{(S)}$ is the necessary number of gates to simulate the free system evolution $U_S(\Delta t)$ in Eq.~\eqref{eqn:evSimTot}, which is efficient under the required assumptions \cite{Lloyd1996}. Consequently, to simulate the dynamics up to time $t$ making an error not bigger than $\epsilon_g$, we need
\begin{equation}
\label{eqn:totNumG}
N_G=\left\lceil ((2R-1)\cdot K \cdot J_k+N_G^{(S)}) f(M)t^2/\epsilon_g\right\rceil
\end{equation}
gates. Under the condition of local GKS operators, we can substitute $R=k$ in Eq.~\eqref{eqn:totNumG}. Once again, $N_G=\textnormal{poly}(M,t,1/\epsilon_g)$ and therefore the MCM is efficiently simulable on a quantum computer according to the dissipative quantum Church-Turing theorem. {This is our third major result}. The total number of gates scales as $t^2/\epsilon_g$, which is optimal \cite{Cleve2017} for collision models. 

\textit{Conclusions}.--We have presented the \textit{multipartite collision model} (MCM), able to reproduce any Markovian dynamics (or, more precisely, any divisible dynamical map) of a general system made of $M$ subsystems by means of elementary interactions between each subsystem and a single environment ancilla, which can be efficiently simulated through elementary quantum gates. Furthermore, we have derived an analytical error bound for the simulation of generic semigroup dynamics via MCM, and observed that it displays an optimal scaling. In light of the above findings, we believe that the MCM will play a major role in the study and simulation of multipartite open quantum systems in the next future. 

%We have shown a simple, easily implementable $1\rightarrow 1$ relation between the coefficients of the most general GKLS master equation and the freely tunable parameters of the MCM, both when the former is in the diagonal and non-diagonal form, the latter being particularly useful in situations in which one is not able to diagonalize an abstract Kossakowski matrix.  In addition, we have estimated an analytical error bound for the simulation of a generic semigroup dynamics by means of the collision model, taking into account any order of the small timestep $\Delta t$, and found that it is optimal for collision models. We have employed such error estimation to show that the MCM is efficiently simulable according to the dissipative quantum Church-Turing theorem. These features make the MCM the most convenient quantum collision model for the study and simulation of general multipartite open quantum systems. 

{Our results pave the way towards general applications of the collision approach to global master equations, many-body dissipative collective effects like superradiance or synchronization, transport in complex open systems, as well as to a wide range of problems in quantum thermodynamics, such as the study of Landauer's principle in any multipartite system, of composed thermal machines or of the microscopic exchange of energy between subsystems and ancillas. Finally, the efficient simulation of the MCM on a NISQ device is within our reach through currently available technology \cite{Garcia-Perez2020a}.}

\textit{Acknowledgments}--
M.C. thanks Rodrigo Mart\'{i}nez-Pe\~{n}a for useful suggestions.
This work was supported by CAIB through QUAREC project (PRD2018/47), by the Spanish State Research Agency through projects PID2019-109094GB-C21, and through the Severo Ochoa and Mar\'ia de Maeztu Program for Centers and Units of Excellence in R\&D (MDM-2017-0711). G.L.G. is funded by the Spanish Ministerio de Educación y Formación Profesional/Ministerio de Universidades and co-funded by the University of the Balearic Islands through the Beatriz Galindo program (BG20 /00085).
G. D. C. acknowledges support from the UK
EPSRC grants EP/S02994X/1 and EP/T026715/1. S.M. acknowledges financial support from the Academy of Finland via the Centre of Excellence program (Project no.~336814). 
\bibliography{draftPRLbiblio}
%\bibliographystyle{vancouver}
%\bibliographystyle{iopart-num}
%\end{multicols}
\newpage
\pagebreak
\widetext
\begin{center}
\textbf{\large Supplemental Material: Collision models can efficiently simulate any multipartite Markovian quantum dynamics}
\end{center}
%%%%%%%%%% Merge with supplemental materials %%%%%%%%%%
%%%%%%%%%% Prefix a "S" to all equations, figures, tables and reset the counter %%%%%%%%%%
\setcounter{equation}{0}
\setcounter{figure}{0}
\setcounter{table}{0}
\setcounter{page}{1}
\makeatletter
\renewcommand{\theequation}{S\arabic{equation}}

\section{Previous collision models for multipartite systems}

In this section we briefly review some previous schemes of collision models focused on multipartite open systems and decomposable into elementary collisions. Let us start considering Refs.  \cite{Daryanoosh2018,DeChiara2020}, where collision models based on elementary subsystem-ancilla interactions with entangled ancillas have been shown to create correlations. The main limitation with respect to our approach is given by the geometrical constraints on the entangled state of the ancillas, which prevents such model from reproducing a general GKLS master equation. For instance, it cannot properly handle the temperature of a bath (see e.g. the examples in Ref.~\cite{Daryanoosh2018}).
\newline \indent  
In Ref.  \cite{Lorenzo2017}, a ``composite collision model'' was introduced; here, the system is partitioned into many auxiliary systems interacting with the environment ancillas. In this model,  the total interaction and the master equations are strictly local, and therefore only  \textit{local} master equations can be  simulated \cite{DeChiara2018a}, as it is not possible to reproduce any global effect \cite{Cattaneo2019b}.

A quite general  \textit{cascade} collision model for correlated quantum channels has been introduced by Giovannetti and Palma \cite{Giovannetti2012} and has been employed in several interesting applications \cite{Lorenzo2015a,Lorenzo2015,Cusumano2017a,Cusumano2018}. The model assumes an ordered set of subsystems where the dynamics of the $m$th subsystem is not influenced by the dynamics of the $m'$th one, with $m'> m$. In general terms, the cascade model cannot reproduce a generic GKLS master equation, unless one applies it several times with permuted series of collisions in order to account for all possible causality dependence, let us say one for each possible order of the set of subsystems, which appears extremely complex and inefficient, both in terms of resources (ancillas and gates) and in terms of readability of the final $1\rightarrow 1$ relation between parameters of the model and GKLS coefficients. Moreover, this cascade model does not include the possibility of non-local unitary system evolutions, and is therefore not suitable for the study of generic global master equations.

\section{Derivation of the collision model and master equation}
\subsection{General derivation of the master equation in the limit of infinitesimal timestep}
We will review here the method to derive the master equation of a collision model with internal unitary dynamics of the system, introduced in Ref.~\cite{Lorenzo2017} or analogously in Ref.~\cite{Landi2014}. For simplicity, let us assume to have a single interaction Hamiltonian $H_I$ describing the system-environment collisions with associated energy constant $g_I$, and the free system Hamiltonian $H_S$ with energy constant $g_S$. Then, the total unitary evolution of the collision model during the timestep $\Delta t$ can be written as ($\hbar=1$ throughout all the supplemental material):
\begin{equation}
\label{eqn:totEv}
U_{sim}(\Delta t)=U_S(\Delta)\circ U_I(\Delta t),\qquad U_S(\Delta t)=e^{-i g_S \Delta t H_S},\qquad U_I(\Delta t)=e^{-i g_I \Delta t H_I}.
\end{equation}
The corresponding quantum map reads:
\begin{equation}
\phi_{\Delta t}[\rho_S]=\Tr_E\left[U_{sim}(\Delta t)\rho_{SE}U_{sim}^\dagger(\Delta t)\right],
\end{equation}
where $\rho_{SE}=\rho_S\otimes\rho_E(0)$.
We set $g_S\ll g_I\ll \Delta t^{-1}$. In this limit, we are able to derive a consistent master equation for $\Delta t\rightarrow 0^+$.
The condition $g_j\Delta t\rightarrow 0^+$ for all $j$ allows us to write a series expansion of $U_{sim}(\Delta t)$: each unitary evolution can be written as (till the second order):
\begin{equation}
U_j(\Delta t)= \mathbb{I}-ig_j\Delta t H_j-\frac{g_j^2\Delta t^2}{2}H_j^2+O(g_j^3\Delta t^3).
\end{equation}
Let us now expand the quantum map $\phi_{\Delta t}$ using the above equation: $\phi_{\Delta t}\simeq\phi_{\Delta t}^{(0)}+\phi_{\Delta t}^{(1)}+\phi_{\Delta t}^{(2)}$, where $\phi_{\Delta t}^{(k)}$ is of the order of $O((g_j\Delta t)^k)$. We obtain the following terms:
\begin{equation}
\phi_{\Delta t}^{(0)}[\rho_S]=\rho_S.
\end{equation}
\begin{equation}
\phi_{\Delta t}^{(1)}[\rho_S]=-i g_S\Delta t [H_S,\rho_S]-i g_I\Delta t \Tr_E[[H_I,\rho_{SE}]].
\end{equation}
\begin{equation}
\begin{split}
\phi_{\Delta t}^{(2)}[\rho_S]=&{g_S^2\Delta t^2\mathcal{D}_{H_S}[\rho_S]}+g_I^2\Delta t^2\Tr_E[\mathcal{D}_{H_I}[\rho_{SE}]]\\
&+{g_Sg_I\Delta t^2\Tr_E[H_S\rho_{SE} H_I+H_I\rho_{SE} H_S-H_SH_I\rho_{SE}-\rho_{SE}H_IH_S]},
\end{split}
\end{equation}
where $\mathcal{D}_{O}[\rho]=O\rho O^\dagger-\frac{1}{2}\{O^\dagger O,\rho\}$. We remove the first-order term proportional to $g_I$ by setting $\Tr_E[[H_I,\rho_{SE}]]=0$, which is the usual requirement in the derivation of the master equation \cite{Cattaneo2019b}. Furthermore, we assume $g_I^2\Delta t \rightarrow \gamma$, where $\gamma$ is a finite constant with the units of energy, while $g_S^j\Delta t\rightarrow 0^+$ for all $j$ (i.e. $g_S=O(\gamma)=O(1)$, having fixed a proper energy scale). This can be analogously achieved, for instance, by setting $g_I=O(g_S)=O(1)$, while $H_I\rightarrow H_I/\sqrt{\Delta t}$ \cite{Landi2014}. Given that $g_I\gg g_S$, we can neglect all the terms of the second order depending on the system Hamiltonian. Finally, the master equation reads:
\begin{equation}
\begin{split}
\label{eqn:compositeMasterEq}
\frac{d\rho_S}{dt}=&\lim_{\Delta t\rightarrow 0^+}\frac{\phi_{\Delta t}[\rho_S]-\rho_S}{\Delta t}=-i g_S [H_S,\rho_S]+\gamma\Tr_E[H_I\rho_{SE} H_I-\frac{1}{2}\{H_I^2,\rho_{SE}\}]+O(\Delta t)+O(g_I \Delta t).
\end{split}
\end{equation}

\subsection{Derivation for the \textit{\textit{multipartite collision model}}}
Consider a pair $p$ of GKS operators $F_{m,\alpha}$, $F_{m',\alpha'}$. Then, the Trotter formula symmetrization \cite{Suzuki1985,Hatano2005} applied to Eq.~\eqref{eqn:pairEv} of the main text leads to:
\begin{equation}
\label{eqn:suzukiSecOrd}
U_p(\Delta t)=U_p^{(m,\alpha)}(\Delta t/2)U_p^{(m',\alpha')}(\Delta t)U_p^{(m,\alpha)}(\Delta t/2)=\exp\{-ig_I\Delta t [(\lambda_p^{(m,\alpha)}F_{m,\alpha}+\lambda_p^{(m',\alpha')}F_{m',\alpha'})\sigma_{p}^++h.c.]\}+O(g_I^3\Delta t^3),
\end{equation}
where the remainder of the above equation can be obtained from the equality \cite{Hatano2005} $e^{g_I\Delta t/2 A}e^{g_I\Delta t/B}e^{g_I\Delta t/2 A}=e^{g_I\Delta t (A+B)+(g_I\Delta t)^3 R_3+(g_I\Delta t)^5 R_5+\ldots}$, where $R_j$ with $j$ odd and $j\geq 3$ are expressions proportional to $j$ operators $A$ and/or $B$, derived from the Baker-Campbell-Hausdorff formula.

Let us now insert all the pairs $p\in\mathsf{P}$ in the product of Eq.~\eqref{eqn:evComposition} of the main text. For simplicity and to recover the notation of the master equation~\eqref{eqn:compositeMasterEq}, we will gather all the terms in a single interaction Hamiltonian written as:
\begin{equation}
\label{eqn:intHamTot}
H_I=\sum_{p\in\mathsf{P}} \left[(\lambda_p^{(m,\alpha)}F_{m,\alpha}+\lambda_p^{(m',\alpha')}F_{m',\alpha'})\sigma_{p}^++h.c.\right],
\end{equation}
and compute the total evolution in Eq.~\eqref{eqn:evComposition} of the main text as $U_I(\Delta t)=\exp(-ig_I\Delta t H_I)$. In general, $\exp(-ig_I\Delta t H_I)\neq\prod_{p\in\mathsf{P}}\exp\{-ig_I\Delta t [(\lambda_p^{(m,\alpha)}F_{m,\alpha}+\lambda_p^{(m',\alpha')}F_{m',\alpha'})\sigma_{p}^++h.c.]\}$, because the system operators of different pairs may not commute, but we will see that this simple $U_I(\Delta t)$ does the job in the master equation up to a suitable order. Let us now derive the master equation in the limit $g_S\ll g_I\ll \Delta t^{-1}$ by inserting Eq.~\eqref{eqn:intHamTot} in Eq.~\eqref{eqn:compositeMasterEq}.
The dissipator $\mathcal{D}$ (term proportional to $\gamma$) reads:
\begin{equation}
\label{eqn:dissipatorOne}
\begin{split}
\mathcal{D}[\rho_S]=&\gamma\sum_{p\in\mathsf{P}}\sum_{p'\in\mathsf{P}}\Bigg[\Tr_E[\sigma_{p'}^-\sigma_{p}^+\rho_E(0)]\left(\left(\lambda_{p}^{(m,\alpha)}F_{m,\alpha}+\lambda_{p}^{(m',\alpha')} F_{m',\alpha'}\right)\rho_S\left((\lambda_{p'}^{(n,\beta)})^*F_{n,\beta}^\dagger+(\lambda_{p'}^{(n',\beta')})^* F_{n',\beta'}^\dagger\right)\right.\\
&\left.-\frac{1}{2}\left\{\left((\lambda_{p'}^{(n,\beta)})^*F_{n,\beta}^\dagger+(\lambda_{p'}^{(n',\beta')})^* F_{n',\beta'}^\dagger\right)\left(\lambda_{p}^{(m,\alpha)}F_{m,\alpha}+\lambda_{p}^{(m',\alpha')}F_{m',\alpha'}\right),\rho_S \right\}\right)\\
&+\Tr_E[\sigma_{p'}^+\sigma_{p}^-\rho_E(0)]\left(\left((\lambda_{p}^{(m,\alpha)})^*F_{m,\alpha}^\dagger+(\lambda_{p}^{(m',\alpha')})^* F_{m',\alpha'}^\dagger\right)\rho_S\left(\lambda_{p'}^{(n,\beta)}F_{n,\beta}+\lambda_{p'}^{(n',\beta')}F_{n',\beta'}\right)\right.\\
&\left.-\frac{1}{2}\left\{\left(\lambda_{p'}^{(n,\beta)}F_{n,\beta}+\lambda_{p'}^{(n',\beta')}F_{n',\beta'}\right)\left((\lambda_{p}^{(m,\alpha)})^*F_{m,\alpha}^\dagger+(\lambda_{p}^{(m',\alpha')})^* F_{m',\alpha'}^\dagger\right),\rho_S \right\}\right)\Bigg],
\end{split}
\end{equation}
where $p=(m,\alpha,m',\alpha')$, $p'=(n,\beta,n',\beta')$, and we have neglected the terms with double creation or double annihilation of an ancillary qubit excitation because of the following initial state choice.

As explained in the main text, we choose as initial state of the ancillas $\rho_E(0)=\bigotimes_{p\in\mathsf{P}} \eta_p$, where $\eta_p=c_p\ket{\downarrow}_p\!\bra{\downarrow}+(1-c_p)\ket{\uparrow}_p\!\bra{\uparrow}$, with $0\leq c_p\leq 1$, is a diagonal state in the basis of $\sigma_{p}^z$. Since $\rho_E(0)$ is a separable state of all the ancillas, the autocorrelation functions of the environment are proportional to the Dirac's deltas $\delta_{m,n}\delta_{m',n'}\delta_{\alpha,\alpha'}\delta_{\alpha',\alpha''}$ (or shortly, $\delta_{p,p'}$ for different pairs) and they can be simplified as $\Tr_E[\sigma_{p'}^-\sigma_{p}^+\rho_E(0)]=c_p\delta_{p,p'}$, $\Tr_E[\sigma_{p'}^+\sigma_{p}^-\rho_E(0)]=(1-c_p)\delta_{p,p'}$. Note that the choice of initial environment state  implies that $\Tr_E[\sigma_{p}^+\rho_E(0)]=0$, that leads to satisfy the requirement $\Tr_E[[H_I,\rho_{SE}]]=0$ necessary to derive Eq.~\eqref{eqn:compositeMasterEq}. More in general, $\Tr_E[\underbrace{\sigma_{p}^{+-}\sigma_{p'}^{+-}\sigma_{p''}^{+-}\ldots}_{j\; (\sigma_E^{+-})'s}\rho_E(0)]=0$ if $j$ is odd. This immediately implies that all the terms proportional to $g_I^j$ with odd $j$ in the master equation vanish. Therefore, the remainder $O(g_I\Delta t)$ in Eq.~\eqref{eqn:compositeMasterEq} vanishes for the MCM, and the leading order is $O(\Delta t)$, or the equivalent $O(g_I^2\Delta t^2)=\gamma O(\Delta t)$. Accordingly, the error that the Suzuki-Trotter symmetrization in Eq.~\eqref{eqn:suzukiSecOrd} brings to the master equation is not of the order of $O(g_I^3\Delta t^3)$, but of $O(g_I^4\Delta t^4)=O(\Delta t^2)$ ($O(\Delta t)$ in Eq.~\eqref{eqn:compositeMasterEq}), and equivalently for the decomposition into $R$ elementary multi-qubit gates in a scenario with many-body GKS operators. Finally, we understand that deriving the master equation by employing $U_I(\Delta t)$ with Hamiltonian in Eq.~\eqref{eqn:intHamTot} instead of the product in Eq.~\eqref{eqn:evComposition} of the main text is accurate up to an error of the order of $O(g_I^4\Delta t^4)=O(\Delta t^2)$ ($O(\Delta t)$ in Eq.~\eqref{eqn:compositeMasterEq}), since GKS operators associated to different ancillas do not mix in the dissipator of Eq.~\eqref{eqn:dissipatorOne}. This is why the quantum map $\phi_{\Delta t}$ simulates $\mathbb{I}+\Delta t\mathcal{L}$ up to an error of the order of $O(\Delta t^2)$, as claimed in Eq.~\eqref{eqn:approxMap} of the main text.

Let us now rewrite the dissipator by highlighting each term, with $p=(m,\alpha,m',\alpha')$:
\begin{equation}
\label{eqn:dissipatorTwo}
\begin{split}
\mathcal{D}[\rho_S]=&\gamma\sum_{p\in\mathsf{P}}\;\sum_{(x,r),(y,s)=(m,\alpha),(m',\alpha')}\left[c_{p}\lambda_{p}^{(x,r)}(\lambda_{p}^{(y,s)})^*\left(F_{x,r} \rho_S F_{y,s}^\dagger-\frac{1}{2}\{F_{y,s}^\dagger F_{x,r},\rho_S\}\right)\right.\\
&\left.+(1-c_p)(\lambda_{p}^{(x,r)})^*\lambda_{p}^{(y,s)}\left(F_{x,r}^\dagger \rho_S F_{y,s}-\frac{1}{2}\{F_{y,s} F_{x,r}^\dagger,\rho_S\}\right)\right].
\end{split}
\end{equation}
Finally, we obtain the master equation as:
\begin{equation}
\label{eqn:masterEqFin}
\begin{split}
\frac{d\rho_S}{dt}=&-i[g_S H_S,\rho_S]+\sum_{p\in\mathsf{P}} \left[\gamma_p^\downarrow \left(F_{m,\alpha}\rho_S F_{m',\alpha'}^\dagger-\frac{1}{2}\{F_{m',\alpha'}^\dagger F_{m,\alpha},\rho_S\}\right)+h.c.\right]\\
&+\sum_{p\in\mathsf{P}}\left[ \gamma_p^\uparrow \left(F_{m,\alpha}^\dagger\rho_S F_{m',\alpha'}-\frac{1}{2}\{F_{m',\alpha'} F_{m,\alpha}^\dagger,\rho_S\}\right)+h.c.\right].
\end{split}
\end{equation}
with
\begin{equation}
\label{eqn:coefficientsBis}
\begin{split}
&\gamma_p^\downarrow=\begin{cases}
\gamma c_p \lambda_p^{(m,\alpha)}(\lambda_p^{(m',\alpha')})^*&\textnormal{ if }m\neq m'\textnormal{ or }\alpha\neq\alpha'\\
\gamma\sum_{\bar{p}} c_{\bar{p}} \abs{\lambda_{\bar{p}}^{(m,\alpha)}}^2 &\textnormal{ otherwise}\\
\end{cases}\\
&\gamma_p^\uparrow=\begin{cases}
\gamma(1-c_p) (\lambda_p^{(m,\alpha)})^*\lambda_p^{(m',\alpha')}&\textnormal{ if }m\neq m'\textnormal{ or }\alpha\neq\alpha'\\
\gamma\sum_{\bar{p}} (1-c_{\bar{p}}) \abs{\lambda_{\bar{p}}^{(m,\alpha)}}^2 &\textnormal{ otherwise},\\
\end{cases}\\
\end{split}
\end{equation}
summing over all the unordered pairs of GKS operators  $\bar{p}=(m,\alpha,\bar{m},\bar{\alpha})$.
 These are Eqs.~\eqref{eqn:Liouvillian} and~\eqref{eqn:coefficients} of the main text. The summation over all the pairs for $m=m'$, $\alpha=\alpha'$ is due to the fact that each interaction $U_p(\Delta t)$ for a pair $p$ of GKS operators also brings a  ``self-contribution'' of the form $F_{m,\alpha}\rho_SF_{m,\alpha}^\dagger+\ldots$ in the dissipator. As can be observed in Eq.~\eqref{eqn:dissipatorTwo}, the ``emission'' contribution brought to the master equation by $U_p(\Delta t)$ for a given pair $p$ of GKS operators can be described by a $2\times 2$ matrix $v_p$, with $(v_p)_{11}=c_p \abs{\lambda_{m,\alpha}}^2$, $(v_p)_{12}=c_p \lambda_{m,\alpha}(\lambda_{m',\alpha'})^*$, $(v_p)_{21}=c_p (\lambda_{m,\alpha})^*\lambda_{m',\alpha'}$, $(v_p)_{22}=c_p \abs{\lambda_{m',\alpha'}}^2$. Clearly, $v_p\geq 0$ since it is obtained by an autocorrelation function. The total Kossakowski matrix for the emission (and correspondingly for absorption), described by the coefficients $\gamma_p^\downarrow$, is obtained by summing all the semipositive matrices $v_p$ (extended as sparse $J\times J$ matrices) for all $p\in\mathsf{P}$, therefore it is semipositive as well and the master equation is in GKLS form.

\section{Remarks on the collision model and examples}
\subsection{Shortcuts}
In case one is not able to diagonalize a symbolic Kossakowski matrix and has to rely on the non-diagonal MCM, some shortcuts may still reduce the number of required resources: while one ancilla for each pair of GKS operators is necessary to control all the degrees of freedom for the most general non-diagonal GKLS master equation, in many situations we do not need such a large number of them.

For instance, let us address the master equation describing the perfectly collective superradiant emission of $M$ identical two-level atoms immersed in a single common bath:
\begin{equation}
\label{eqn:supMasEq}
\frac{d}{dt}\rho_S(t)=-i[H_A+H_{LS},\rho_S(t)]+\sum_{m,m'=1}^M \gamma\left(\sigma_m^-\rho_S(t)\sigma_{m'}^+-\frac{1}{2}\{\sigma_{m'}^+\sigma_m^-,\rho_S(t)\}\right),
\end{equation}
with $H_A=\sum_{m=1}^M \frac{\omega}{2}\sigma_m^z$, and $H_{LS}=\sum_{m,m'=1}^M s^\downarrow\sigma_{m'}^+\sigma_m^-$ is the Lamb-shift Hamiltonian, which can be simulated by means of two-qubit gates. Then, we can employ a version of the MCM requiring only a single ancilla in the ground state, that couples to each atom through $\sigma_E^+$: let us build the elementary collisions through the two-qubit gates $U^{(m)}(\Delta t)=\exp\{-i g_I\Delta t (\sigma_m^-\sigma_E^++h.c.)\}$, corresponding to Eq.~\eqref{eqn:elemInt} of the main text. Then, we can build the symmetrized unitary operator (with a total number of $2M-1$ gates):
\begin{equation}
\label{eqn:supU}
U_{sup}(\Delta t)=\prod_{m=M}^1U^{(m)}(\Delta t/2)\prod_{m'=1}^MU^{(m')}(\Delta t/2),
\end{equation}   
corresponding to $U_p(\Delta t)$ in Eq.~\eqref{eqn:pairEv} of the main text. By running the MCM with $U_I(\Delta t)=U_{sup}(\Delta t)$, $U_S(\Delta t)=\exp(-i g_S\Delta t(H_A+H_{LS})/g_S)$ as in Eq.~\eqref{eqn:totEv} (we assume $\omega=O(g_S)$), with the usual requirement $g_S\ll g_I\ll \Delta t^{-1}$, $g_I^2\Delta t\rightarrow\gamma$, we readily get the master equation~\eqref{eqn:supMasEq}. The same result can be obtained in the case of an inhomogeneous spatial distribution of the atoms, i.e. when the decay rate $\gamma$ is not uniform anymore and we have to describe it through a suitable Kossakowksi matrix $\gamma_{mm'}$. In this scenario, we need to introduce a proper weight $\lambda^{(m)}=\xi_m$ in each $U^{(m)}(\Delta t)$. For instance, if each atom couples to the electromagnetic field with a given dimensionless strength $\xi_m$, we will have $\gamma_{mm'}=\gamma\xi_m\xi_{m'}$, and correspondigly we will need to set $U^{(m)}(\Delta t)=\exp\{-i g_I\Delta t (\xi_m\sigma_m^-\sigma_E^++h.c.)\}$.

In contrast, in a situation with a chain of $M$ strongly coupled two-level atoms, each one immersed in a local bath, we cannot rely on a single ancilla. Nonetheless, we can describe the master equation of such scenario as a superposition of $M$ common baths, each one bringing a term as in Eq.~\eqref{eqn:supMasEq} (see for instance Ref.~\cite{Cattaneo2020}). Then, we can select one ancilla for each different bath, and implement the chain of elementary two-qubit gates as in Eq.~\eqref{eqn:supU}, for each of them. We thus obtain a further shortcut of the MCM for the non-diagonal GKLS master equation, which is able to simulate the open dynamics of the atomic chain by employing $M$ ancillas only.

\subsection{Temperature}
For simplicity, let us consider a common thermal bath of harmonic oscillators collectively acting on two identical qubits with frequency $\omega$, and let us suppose that the bath is at temperature $T$. Then, similarly to Eq.~\eqref{eqn:supMasEq}, the master equation reads:
\begin{equation}
\label{eqn:tempEq}
\begin{split}
\frac{d}{dt}\rho_S(t)=&-i[H_q+H_{LS},\rho_S(t)]+\underbrace{\sum_{m,m'=1,2} \gamma^\downarrow\left(\sigma_m^-\rho_S(t)\sigma_{m'}^+-\frac{1}{2}\{\sigma_{m'}^+\sigma_m^-,\rho_S(t)\}\right)}_\textnormal{emission}\\
&+\underbrace{\sum_{m,m'=1,2} \gamma^\uparrow\left(\sigma_m^+\rho_S(t)\sigma_{m'}^--\frac{1}{2}\{\sigma_{m'}^-\sigma_m^+,\rho_S(t)\}\right)}_\textnormal{absorption},
\end{split}
\end{equation}
where $H_q=\omega/2(\sigma_1^z+\sigma_2^z)$ is the free Hamiltonian of the qubits, $H_{LS}$ is the Lamb-shift Hamiltonian and the emission and absorption coefficients are given by:
\begin{equation}
\label{eqn:emisAbs}
\gamma^\downarrow=\gamma (N_T(\omega)+1),\qquad \gamma^\uparrow=\gamma N_T(\omega), 
\end{equation}
where $N_T(\omega)=1/(\exp(\beta\omega)-1)$, with $\beta=1/k_B T$. Then, the MCM still requires a single ancillary qubit to simulate the master equation~\eqref{eqn:tempEq}, following the same lines of the previous shortcut. Indeed, let us consider a single ancilla with operator $\sigma_E^+$, initialized in the state $\eta_E=c_E\ket{\downarrow}_E\!\bra{\downarrow}+(1-c_E)\ket{\uparrow}_E\!\bra{\uparrow}$, with $c_E=(N_T(\omega)+1)/(2N_T(\omega)+1)$. Then, we build the elementary collision gates as $U^{(m)}(\Delta t)=\exp\{-ig_I\Delta t \lambda( \sigma_m^-\sigma_E^++h.c.)\}$, where $\lambda=\sqrt{2N_T(\omega)+1}$. The total evolution for the system-environment interaction is then built as (equivalent to Eqs.~\eqref{eqn:pairEv} and~\eqref{eqn:evComposition} of the main text):
\begin{equation}
\label{eqn:totalIntEv}
U_I(\Delta t)=U^{(1)}(\Delta t/2)U^{(2)}(\Delta t)U^{(1)}(\Delta t/2),
\end{equation}
and following the derivation of the MCM we straightforwardly obtain the master equation~\eqref{eqn:tempEq}. Note that the proper engineering of the parameter $\lambda$ in each elementary interaction plays here a central role: if the temperature increases, for instance, we just have to tune the value of $\lambda$ and to suitably change the temperature of the initial ancillary state. Through this procedure, we can simulate any temperature of even more complex baths.

\section{Error bound derivation}
For simplicity, let us derive the error bound by assuming the $k$-locality condition, i.e. $\mathcal{L}=\sum_{\sigma=1}^K\mathcal{L}_\sigma$ and each $\mathcal{L}_\sigma$ acts non-trivially on $k$ subsystems only. Each $\mathcal{L}_\sigma$ has at maximum $J_k=d^{2k}-1$ Lindblad operators. The result without $k$-locality can then be recovered by setting $k=M$, $K=1$, $J_k=J$. First of all, let us estimate the maximum number $K$ of possible different $k$-local Liouvillians in the presence of $M$ subsystems. We can evaluate it as the number of $k$-element combinations of $M$ objects, for simplicity without repetition (Liouvillian terms with repetitions can be merged with larger terms without repetition):
\begin{equation}
\label{eqn:K}
K=\frac{M!}{k!(M-k)!}.
\end{equation} 
To perform an accurate resource analysis for quantum computation, it is important to study the behavior of $K$ in the asymptotic limit $M\rightarrow \infty$. Using Stirling's formula, we have for fixed $k$ and $M\rightarrow \infty$:
\begin{equation}
\begin{split}
\frac{M!}{k!(M-k)!}&\sim\frac{\sqrt{2\pi M}M^M e^{-M}}{k!\sqrt{2\pi (M-k)}(M-k)^{(M-k)} e^{-(M-k)}}\sim\frac{M^k}{k!e^k},
\end{split}
\end{equation}
showing that it behaves polynomially as a function of the number of subsystems (note that we obtain the same asymptoptic behavior if we assume combinations with repetitions). Furthermore, for fixed $k$ and $M$ we always have fewer combinations than variations, therefore $K\leq M^k$ for all $M,k$.

As discussed in the main text, we employ the $1\rightarrow 1$ superoperator norm to estimate the error bound: $\norm{\mathcal{T}}_{1\rightarrow 1}=\sup_{\norm{A}_1=1}\norm{\mathcal{T}[A]}_1$, where $\norm{A}_1=\Tr[\sqrt{A^\dagger A}]$ is the trace norm. As noticed in Ref.~\cite{Kliesch2011}, this norm does not behave well as the system size increases, since $\norm{A\otimes\mathbb{I}_n}_1=n\norm{A}_1$, nonetheless we can reduce it to the computation of some operator norms by means of Hölder's inequality: $\norm{AB}_1\leq\norm{A}_1\norm{B}_\infty$. $\norm{A}_\infty=\sup_{\norm{v}=1}\norm{Av}$ is the operator or infinity norm, where $\norm{v}$ is the standard vector norm. The superoperator norm, the trace norm and the operator norm satisfy the submultiplicativity property. Let us now evaluate an error bound for the truncation error.

\subsection{Truncation error}
We want to evalute the truncation error $\epsilon_t=\norm{\exp\mathcal{L}\Delta t-(\mathbb{I}+\Delta t\mathcal{L})}_{1\rightarrow 1}$ for the simulation of a generic Liouvillian $\mathcal{L}$. This is basically the second-order remainder of the Taylor expansion of $\exp\mathcal{L}t$ around zero. Taking inspiration from Ref.~\cite{Childs2018a} (Supp. Inf.), we have:
\begin{equation}
\begin{split}
\epsilon_t&=\norm{\exp\mathcal{L}\Delta t-(\mathbb{I}+\Delta t \mathcal{L})}_{1\rightarrow 1}=\norm{\sum_{j=2}^\infty \frac{(\mathcal{L}\Delta t)^j}{j!}}_{1\rightarrow 1}\leq \sum_{j=2}^\infty \frac{(\norm{\mathcal{L}}_{1\rightarrow 1}\Delta t)^j}{j!}\leq \frac{(\norm{\mathcal{L}}_{1\rightarrow 1}\Delta t)^2}{2!}\sum_{j=0}^\infty \frac{(\norm{\mathcal{L}}_{1\rightarrow 1}\Delta t)^j}{j!},
\end{split}
\end{equation}
therefore we get:
\begin{equation}
\boxed{\epsilon_t\leq \frac{(\norm{\mathcal{L}}_{1\rightarrow 1}\Delta t)^2}{2}\exp(\norm{\mathcal{L}}_{1\rightarrow 1}\Delta t)\leq  e\frac{(\norm{\mathcal{L}}_{1\rightarrow 1}\Delta t)^2}{2},}
\end{equation}
with the prescription $\norm{\mathcal{L}}_{1\rightarrow 1}\Delta t<1$ as in Ref.~\cite{Berry2007}.

Let us finally estimate $\norm{\mathcal{L}}_{1\rightarrow 1}$ for a $k$-local Liouvillian. Using the decomposition $\mathcal{L}=\sum_{\sigma=1}^K\mathcal{L}_\sigma$, we have $\norm{\mathcal{L}}_{1\rightarrow 1}\leq K \norm{\mathcal{L}_{max}}_{1\rightarrow 1}$, where $\norm{\mathcal{L}_{max}}_{1\rightarrow 1}=\max_\sigma\norm{\mathcal{L}_\sigma}_{1\rightarrow 1}$. Since any Liouvillian $\mathcal{L}_\sigma$ is $k$-local, it can be seen as a sparse matrix \cite{Childs2017} acting on $k$ elements only (out of $M$). The maximal norm does not increase when the number of subsystems increases, therefore the truncation error has a polynomial dependence on $M$ (the exponential does not provide an ``exploding'' behaviour, because in general we set the condition $\norm{\mathcal{L}}_{1\rightarrow 1}\Delta t< 1$). We can further estimate $\norm{\mathcal{L}}_{1\rightarrow 1}$ following the original work on $k$-local Liouvillians \cite{Kliesch2011}, and find:
\begin{equation}
\begin{split}
\norm{\mathcal{L}_\sigma}_{1\rightarrow 1}&=\sup_{\norm{\rho}_1=1}\norm{-i[H_\sigma,\rho]+\sum_{j=1}^{J_k} \left(L_{\sigma, j}\rho L_{\sigma,j} ^\dagger-\frac{1}{2}\{L_{\sigma,j} ^\dagger L_{\sigma,j},\rho\}\right)}_{1}\\
&\leq 2\norm{H_\sigma}_\infty+2\sum_{j=1}^{J_k}\norm{L_{\sigma,j}}_\infty^2\leq 2(a_\sigma+J_k a_\sigma^2),
\end{split}
\end{equation}
with $a_\sigma=\max_j\{\norm{H_\sigma}_\infty,\norm{L_{\sigma,j}}_\infty\}$, where we have used the Hölder's inequality and the fact that $\norm{A}_\infty=\norm{A^\dagger}_\infty$. Consequently, we have:
\begin{equation}
\label{eqn:truncErrOne}
\norm{\mathcal{L}}_{1\rightarrow 1}\leq K\norm{\mathcal{L}_{max}}_{1\rightarrow 1}\leq 2K(a_{max}+J_k a_{max}^2),
\end{equation}
with $a_{max}=\max_\sigma a_\sigma$. Being a supremum over operator norms of bounded operators, $a_{max}$ does not increase with the number of subsystems. Finally, for simplicity we can write the truncation error bound as:
\begin{equation}
\label{eqn:truncErr}
\boxed {\epsilon_t\leq 2e (K(a_{max}+J_k a_{max}^2)\Delta t)^2,}
\end{equation}
with the prescription $2K(a_{max}+J_k a_{max}^2)\Delta t<1$ \cite{Berry2007}. We will see later how this expression is related to the result given in Eq.~\eqref{eqn:boundT} of the main text. 

Note that the derivation of a bound for the truncation error is not restricted to the MCM, but \textit{is valid for any collision model} simulating the semigroup dynamics driven by $\mathcal{L}$.

\subsection{Collision error}
Let us finally estimate an error bound for the collision error $\epsilon_c=\norm{\phi_{\Delta t}-(\mathbb{I}+\Delta t\mathcal{L})}_{1\rightarrow 1}$, where $\phi_{\Delta t}$ is the MCM quantum map for the simulation of a generic Liouvillian $\mathcal{L}$. We will find that its expression is more complex than the one of the truncation error. For simplicity, we will evaluate it for the MCM for a GKLS master equation in diagonal form, and then discuss the differences with the non-diagonal form. In contrast, we will not assume the locality of the GKS operators, which may be many-body as well.

Let us start with defining the form of the global unitary transformation evolving the collision model (Eq.~\eqref{eqn:evSimTot} of the main text):
\begin{equation}
\label{eqn:decompositionSim}
U_{sim}(\Delta t)=\exp(-i\Delta t g_S H_S)\prod_{\sigma=1}^K\prod_{j=1}^{J_k}\prod_{r=R}^{1} \exp(-ig_I\Delta t/2\mu_r^{(\sigma,j)} H_r^{(\sigma,j)})\prod_{r'=1}^{R} \exp(-ig_I\Delta t/2\mu_{r'}^{(\sigma,j)} H_{r'}^{(\sigma,j)}),
\end{equation}
where we recall that $\sigma$ labels the different $k$-local Liouvillians, $j$ labels the Lindblad operators inside each Liouvillian and $r$ labels the decomposition of each Lindblad operator into elementary quantum gates available in the lab: $L_{\sigma,j}\sigma_{E_{\sigma,j}}^++h.c.=\sum_{r=1}^R\mu_r^{(\sigma,j)} H_r^{(\sigma,j)}$ (see the related comment about extensions to non-local GKS operators in the main text).  The collision error reads:
\begin{equation}
\begin{split}
\epsilon_c=\sup_{\norm{\rho_S}_1=1}\norm{\Tr_E[U_{sim}(\Delta t)\rho_S\otimes\rho_E U_{sim}^\dagger(\Delta t)]-(\mathbb{I}+\Delta t \mathcal{L})}_1=\sup_{\norm{\rho_S}_1=1}\norm{\mathsf{R}_c\left(\Tr_E[U_{sim}(\Delta t)\rho_S\otimes\rho_E U_{sim}^\dagger(\Delta t)]\right)}_1,
\end{split}
\end{equation}
where $\mathsf{R}_c\left(\Tr_E[U_{sim}(\Delta t)\rho_S\otimes\rho_E U_{sim}^\dagger(\Delta t)]\right)$ indicates the remainder of the expansion in Eq.~\eqref{eqn:compositeMasterEq} that is not cancelled out by the substraction with the first order expansion of the Lindblad semigroup. Since the remainder consists of an infinite sum of terms and the partial trace is a linear operation, according to Ref.~\cite{Lidar2008} we can remove the latter:
\begin{equation}
\begin{split}
\norm{\mathsf{R}_c\left(\Tr_E[U_{sim}(\Delta t)\rho_S\otimes\rho_E U_{sim}^\dagger(\Delta t)]\right)}_1&=\norm{\Tr_E\left[\mathsf{R}_c\left(U_{sim}(\Delta t)\rho_S\otimes\rho_E U_{sim}^\dagger(\Delta t)\right)\right]}_1\\
&\leq \norm{\mathsf{R}_c\left(U_{sim}(\Delta t)\rho_S\otimes\rho_E U_{sim}^\dagger(\Delta t)\right)}_1.
\end{split}
\end{equation}
Then, extending a method for Hamiltonian simulation to the framework of open systems \cite{Childs2018a}, we expand the unitary evolutions:
\begin{equation}
\label{eqn:expansionGen}
\begin{split}
&\norm{\mathsf{R}_c\left(U_{sim}(\Delta t)\rho_S\otimes\rho_E U_{sim}^\dagger(\Delta t)\right)}_1=\Biggl|\!\Biggl|\sideset{}{'}\sum \prod_{\sigma=1}^K\prod_{j=1}^{J_k}\prod_{r=R}^{1}\prod_{r'=1}^{R}\\
&\frac{(-i \Delta t g_s H_S)^{w_s}}{w_s!}\frac{(-i \Delta t/2 g_I\mu_r^{(\sigma,j)} H_r^{(\sigma,j)})^{w_{i(\sigma,j,r)}}}{w_{i(\sigma,j,r)}!}\frac{(-i \Delta t/2 g_I\mu_{r'}^{(\sigma,j)} H_{r'}^{(\sigma,j)})^{w'_{i(\sigma,j,r')}}}{w'_{i(\sigma,j,r')}!}\rho_S\otimes\rho_E\\
&\frac{(i \Delta t/2 g_I\mu_{r'}^{(\sigma,j)} H_{r'}^{(\sigma,j)})^{u'_{i(\sigma,j,r')}}}{u'_{i(\sigma,j,r')}!}\frac{(i \Delta t/2 g_I\mu_r^{(\sigma,j)} H_r^{(\sigma,j)})^{u_{i(\sigma,j,r)}}}{u_{i(\sigma,j,r)}!}\frac{(i \Delta t g_s H_S)^{u_{s}}}{u_{s}!}\Biggl|\!\Biggl|_1\\
&\leq\sideset{}{'}\sum\prod_{\sigma=1}^K\prod_{j=1}^{J_k}\prod_{r=R}^{1}\prod_{r'=1}^{R}\\
&\frac{( \Delta t g_S \norm{H_S}_\infty)^{w_s}}{w_s!}\frac{( \Delta t/2 g_I \norm{\mu_r^{(\sigma,j)} H_r^{(\sigma,j)}}_\infty)^{w_{i(\sigma,j,r)}}}{w_{i(\sigma,j,r)}!}\frac{( \Delta t/2 g_I \norm{\mu_{r'}^{(\sigma,j)} H_{r'}^{(\sigma,j)}}_\infty)^{w'_{i(\sigma,j,r')}}}{w'_{i(\sigma,j,r')}!}\norm{\rho_S\otimes\rho_E}_1\\
&\frac{(\Delta t/2 g_I \norm{\mu_{r'}^{(\sigma,j)} H_{r'}^{(\sigma,j)}}_\infty)^{u'_{i(\sigma,j,r')}}}{u'_{i(\sigma,j,r')}!}\frac{( \Delta t/2 g_I \norm{\mu_{r}^{(\sigma,j)} H_{r}^{(\sigma,j)}}_\infty)^{u_{i(\sigma,j,r)}}}{u_{i(\sigma,j,r)}!}\frac{( \Delta t g_S \norm{H_S}_\infty)^{u_{s}}}{u_{s}!}\\
&\leq \mathsf{R}_c\left(\exp{2\Delta t\left[g_S\sum_{\sigma}\norm{H_S^{(\sigma)}}_\infty+\sum_{\sigma,j,r} g_I \norm{\mu_{r}^{(\sigma,j)} H_{r}^{(\sigma,j)}}_\infty\right]}\right).
\end{split}
\end{equation}
We have introduced the notation $\sideset{}{'}\sum $ to express the sum over all the indexes $w_{s},u_{s},w_{i(\sigma,j,r)},w'_{i(\sigma,j,r')},u_{i(\sigma,j,r)},u'_{i(\sigma,j,r)}=1,\ldots,\infty$, such that their possible combinations in the expansion of Eq.~\eqref{eqn:expansionGen} are contained in the remainder $\mathsf{R}_c$. Then, we have employed the Hölder's inequality, the submultiplicativity of the norm, the triangle inequality and the fact that \cite{Lidar2008} $\norm{\rho_S\otimes\rho_E}_1=\norm{\rho_S}_1\norm{\rho_E}_1=1$, since $\norm{\rho_S}_1=1$ by definition in the supremum and $\norm{\rho_E}_1=1$ because it is a density matrix. Furthermore, we have decomposed the $k$-local system Hamiltonian as $H_S=\sum_{\sigma=1}^K H_S^{(\sigma)}$. 

Finally, let us simplify the result by extracting a maximal value: $\Lambda=\max_{\sigma,j,r}\{\norm{H_S^{\sigma}}_\infty,\norm{\mu_{r}^{(\sigma,j)} H_{r}^{(\sigma,j)}}_\infty\}$. Then, we have:
\begin{equation}
\label{eqn:collBis}
\boxed{\epsilon_c\leq \mathsf{R}_c\left(\exp{2\Delta t\Lambda(Kg_S+ \Xi g_I)}\right),}
\end{equation}
where $\Xi=K\cdot J_k\cdot R$. If we employed the MCM for the non-diagonal GKLS form, we would find a product over all the pairs $p\in\mathsf{P}_\sigma$ instead of over all the Lindblad operators for $j=1,\ldots,J_k$ in Eq.~\eqref{eqn:decompositionSim}. Then, we would obtain the same result of Eq.~\eqref{eqn:collBis} with $\Lambda$ maximized over $p\in\mathsf{P}_\sigma$ and $\Xi=K\cdot\abs{\mathsf{P}}_k\cdot R$ ($\abs{\mathsf{P}}_k=\max_\sigma \abs{\mathsf{P}}_\sigma)$. Without assuming the $k$-locality condition, we have the same results without the maximization over $\sigma$, and respectively $\Xi=J\cdot R$ or $\Xi=R\cdot \abs{\mathsf{P}}$, which are the quantities given in the main text. Note that the estimation of the error bound in Eq.~\eqref{eqn:collBis} is once again valid \textit{for any collision model}, which is in general driven by a unitary operator as the one decomposed in Eq.~\eqref{eqn:decompositionSim} and whose quantum map leads to the expansion treated in Eq.~\eqref{eqn:expansionGen}. The only difference between diverse collision models relies on the number of elementary gates appearing in the decomposition of Eq.~\eqref{eqn:decompositionSim}, expressed by the constant $\Xi$, and consequently in the maximization through which we obtain $\Lambda$.

Finally, let us connect the constant $\Lambda$ to the constant $a_{max}$ introduced in Eq.~\eqref{eqn:truncErrOne} for the truncation error. $a_{max}$ is obtained through a maximization over the Lindblad operators, while $\Lambda$ over all the elementary Hamiltonians $\mu_{r}^{(\sigma,j)} H_{r}^{(\sigma,j)}$ providing the decomposition of each Lindblad operator as $L_{\sigma,j}\sigma_{E_{\sigma,j}}^++h.c.=\sum_{r=1}^R\mu_{r}^{(\sigma,j)} H_{r}^{(\sigma,j)}$. Then, we can write $a_{max}\leq R\cdot\Lambda$, having used $\norm{\sigma_{E_{\sigma,j}}^+}_\infty=1$ and $\norm{A}_\infty=\norm{A\otimes\sigma^++A^\dagger\otimes\sigma^-}_\infty$. We can prove the latter result by noticing that $\norm{(A\otimes\sigma^++A^\dagger\otimes\sigma^-)^2}_\infty=\norm{(A\otimes\sigma^++A^\dagger\otimes\sigma^-)}_\infty^2=\norm{AA^\dagger\otimes\ket{\uparrow}\bra{\uparrow}+A^\dagger A\otimes\ket{\downarrow}\!\bra{\uparrow}}_\infty=\max\{\norm{AA^\dagger}_\infty,\norm{A^\dagger A}_\infty\}=\norm{A}_\infty^2$, where we have used $\norm{A A^\dagger}_\infty=\norm{A}_\infty^2$. Consequently, we can rewrite the error bound for the truncation error Eq.~\eqref{eqn:truncErr} as:
\begin{equation}
\label{eqn:truncFin}
\boxed {\epsilon_t\leq 2e (KR\Lambda(1+J_k R\Lambda)\Delta t)^2,}
\end{equation}
with $2KR\Lambda(1+J_k R\Lambda)\Delta t<1$, corresponding to Eq.~\eqref{eqn:boundT} of the main text for $K=1$, $J_k=J$.

Let us now evaluate an explicit expression for the remainder $\mathsf{R}_c$ according to the discussion in the derivation of the general master equation~\eqref{eqn:compositeMasterEq}. We can recognize three different expansions of $\mathsf{R}_c$: one containing only terms with $g_I$, one containing only terms with $g_S$ and one containing both $g_I$ and $g_S$. Let us write them as:
\begin{equation}
\begin{split}
\mathsf{R}_c^{(I)}=&\sum_{i=3}^\infty \frac{(2\Xi\Delta t g_I\Lambda)^i}{i!}\leq\frac{(2\Xi\Delta t g_I\Lambda)^3}{3!}\exp(2\Xi\Delta t g_I\Lambda)=\frac{(2\Xi\Lambda)^3\gamma\Delta t^2g_I}{3!}\exp(2\Xi\Delta t g_I\Lambda),\\
\mathsf{R}_c^{(S)}=&\sum_{i=2}^\infty \frac{(2K\Delta t g_S\Lambda)^i}{i!}\leq\frac{(2K\Delta t g_S\Lambda)^2}{2!}\exp(2R\Delta t g_S\Lambda),\\
\mathsf{R}_c^{(SI)}=&\sum_{i=2}^\infty\frac{(2\Delta t \Lambda)^i\sum_{i'=1}^{i-1}\binom{i}{i'}(Kg_S)^{i'}(\Xi g_I)^{i-i'}}{i!}\leq\sum_{i=2}^\infty\frac{(4\Delta t \Lambda)^iKg_S(\Xi g_I)^{i-1}}{i!}\\
\leq & \frac{(4\Lambda)^2 Kg_S \Xi\Delta t^2 g_I}{2!}\exp(4\Xi\Delta t g_I\Lambda),\\
\end{split}
\end{equation}
where, as discussed in the derivation of the master equation for a generic collision model, $\gamma=\Delta t g_I^2$, and for the last remainder we have used $g_S\ll g_I$ and $\sum_{i'=1}^{i-1}\binom{i}{i'}\leq\sum_{i'=0}^{i}\binom{i}{i'}=2^i$. Finally, we get the result:
\begin{equation}
\label{eqn:collErr}
\boxed{\epsilon_c\leq \frac{(2\Xi\Lambda)^3\gamma\Delta t^2g_I}{3!}\exp(2\Xi\Delta t g_I\Lambda)+\frac{(2 K g_S\Lambda)^2\Delta t^2}{2!}\exp(2\Delta t Kg_S\Lambda)+ \frac{(4\Lambda)^2 Kg_S\Xi \Delta t^2 g_I}{2!}\exp(4\Xi\Delta t g_I\Lambda).}
\end{equation}

The error bound above is valid \textit{for any collision model} derived through the method discussed at the beginning of the Supplemental Material. Note that the expression in Eq.~\eqref{eqn:collErr} does not contain any exponential function in the number of subsystems, therefore the collision model is efficiently simulable according to the dissipative quantum Church-Turing theorem \cite{Kliesch2011}. However, it is suboptimal in the dependence on $\Delta t$: the first and the third term on the right-hand side of Eq.~\eqref{eqn:collErr} scale as $O(\Delta t^{3/2})$, having used $g_I=\sqrt{\gamma/\Delta t}$. As a consequence, the global error shows a behavior (see the discussion in the main text) $\epsilon_g=O(n\Delta t^{3/2})=O(t^{3/2}/\sqrt{n})$, and the number of repetitions of the algorithm is of the order of $n=O(t^3/\epsilon_g^2)$. As discussed in Ref.~\cite{Cleve2017}, ``the best we can do'' in a collision model is likely to have a scaling of $\epsilon_g=O(n\Delta t^2)$. We will now show that such scaling is recovered when choosing the \textit{\textit{multipartite collision model}}.

From the discussion on the derivation of the MCM leading to Eq.~\eqref{eqn:dissipatorTwo}, if we take as bath operators $\sigma_E^{+-}$ and as initial environment state a density matrix diagonal in the computational basis, then we observe that all the terms in the master equation with an odd number of $g_I$'s are removed by the partial trace, and we should not consider them in the above calculations. This means that the remainders are modified as follows:
\begin{equation}
\begin{split}
\mathsf{R}_c^{(I)}=&\sum_{i=2}^\infty \frac{(2\Xi\Delta t g_I\Lambda)^{2i}}{(2i)!}\leq\frac{(2\Xi\Delta t g_I\Lambda)^4}{4!}\cosh(2\Xi\Delta t g_I\Lambda)=\frac{(2\Xi\Lambda)^4\gamma^2\Delta t^2}{4!}\cosh(2\Xi\Delta t g_I\Lambda),\\
\mathsf{R}_c^{(SI)}=&\sum_{i=1}^\infty\frac{(2\Delta t \Lambda)^{2i+1}\sum_{i'=1}^{i}\binom{2i+1}{2i'}(Kg_S)^{2i-2i'+1}(\Xi g_I)^{2i'}}{(2i+1)!}+\sum_{i=2}^\infty\frac{(2\Delta t \Lambda)^{2i}\sum_{i'=1}^{i-1}\binom{2i}{2i'}(Kg_S)^{2i-2i'}(\Xi g_I)^{2i'}}{(2i)!}\\
\leq& Kg_S\sum_{i=1}^\infty\frac{(4\Delta t \Lambda)^{2i+1}(\Xi g_I)^{2i}}{(2i+1)!}+(Kg_S)^2\sum_{i=2}^\infty\frac{(4\Delta t \Lambda)^{2i}(\Xi g_I)^{2i-2}}{(2i)!}\\
\leq& \frac{(4\Lambda)^3 \Xi^2 g_S K\gamma \Delta t^2}{2!}\sinh(4\Xi\Delta t g_I\Lambda)+\frac{(4\Lambda)^4 \Xi^2 (Kg_S)^2 \gamma \Delta t^3}{4!}\cosh(4\Xi\Delta t g_I\Lambda),\\
\end{split}
\end{equation}   
where for simplicity we have approximated the partial binomial coefficient through the total binomial coefficient (tighter approximations may be found). We observe that we have recovered the desired limit of $\epsilon_c=O(\Delta t^2)$, corresponding to the best upper bound for the error of a collision model: for the global error, we have $\epsilon_g=O(n\Delta t^2)$, as claimed in Eq.~\eqref{eqn:globalErrBeh} of the main text. Let us  finally provide a simplified expression for the error bound, by requiring $4\Xi\Delta t g_I\Lambda<1$, $2\Delta t Kg_S\Lambda<1$. This requirement is polynomial in the trade-off between $\Xi$ and $K$ (increasing with the number of subsystems) and $\Delta t$, therefore the simulation is still efficient. We have:
\begin{equation}
\label{eqn:collErrFin}
\boxed{\epsilon_c\leq \textnormal{pol}_1(\Lambda,\Xi,K,g_S,\gamma)\Delta t^2+\textnormal{pol}_2(\Lambda,\Xi,K,g_S,\gamma)\Delta t^3,}
\end{equation} 
with
\begin{equation}
\label{eqn:pol1pol2}
\begin{split}
&\boxed{\textnormal{pol}_1(\Lambda,\Xi,K,g_S,\gamma)= \frac{(2\Xi\Lambda)^4\gamma^2\cosh(1/2)}{4!}+\frac{(4\Lambda)^3 \Xi^2 g_S K\gamma\sinh(1)}{2!}+\frac{e(2 K g_S\Lambda)^2}{2!},}\\
&\boxed{\textnormal{pol}_2(\Lambda,\Xi,K,g_S,\gamma)= \frac{(4\Lambda)^4 \Xi^2 (Kg_S)^2 \gamma \cosh(1)}{4!},}
\end{split}
\end{equation}
that without assuming the $k$-locality condition ($K=1$) correspond to the functions introduced in Eq.~\eqref{eqn:boundC} of the main text.

Finally, let us estimate the polynomial function $f(M)$ appearing in the resource estimation in Eq.~\eqref{eqn:totNumG} of the main text:
\begin{equation}
\label{eqn:fM}
\boxed{f(M)=\frac{(2\Xi\Lambda)^4\gamma^2\cosh(1/2)}{4!}+\frac{(4\Lambda)^3 \Xi^2 g_S K\gamma\sinh(1)}{2!}+\frac{e(2 K g_S\Lambda)^2}{2!}+\frac{(4\Lambda)^4 \Xi^2 (Kg_S)^2 \gamma\Delta t \cosh(1)}{4!}+2e(KR\Lambda(1+J_k R\Lambda))^2.}
\end{equation}
The dependence on $M$ appears in $K$, as prescribed by Eq.~\eqref{eqn:K}, and in $\Xi=K\cdot J_k\cdot R$. Recalling that $K\leq M^k$, $f(M)$ is a polynomial function of the number of subsystems. 

\end{document}